\documentclass[11pt]{article}
\usepackage[T2A]{fontenc}
\usepackage[utf8]{inputenc}
\usepackage{amssymb,amsmath,amsfonts,amsthm,latexsym}

\usepackage{amsfonts,amsmath}
\usepackage{graphicx}
\usepackage{graphics}
\usepackage{caption,subcaption}

\newtheorem{conclusion}{Numerical Conclusion}

\def\R{{\mathbb R}}
\def\C{{\mathbb C}}

\begin{document}

\title{On numerical study of the discrete spectrum of a two-dimensional Schr\"odinger operator with soliton potential}
\author{A.N. Adilkhanov \thanks{National Laboratory ``Astana'', Nazarbayev University, 53, Kabanbay batyr Ave., Astana, 010000, Republic of Kazakhstan; e-mail: aadilkhanov@nu.edu.kz} \and I.A. Taimanov \thanks{Sobolev Institute of Mathematics, 630090 Novosibirsk, Russia, and Novosibirsk State University,
630090 Novosibirsk, Russia; e-mail: taimanov@math.nsc.ru}}
\date{}
\maketitle

\begin{abstract}
The discrete spectra of certain two-dimensional Schr\"odinger operators are numerically calculated. These operators are obtained by the Moutard transformation and have
interesting spectral properties: their kernels are multi-dimensional and the deformations of potentials via the Novikov--Veselov equation
(a two-dimensional generalization of the Korteweg--de Vries equation) lead to blowups. The calculations supply the numerical evidence for some statements about the integrable systems related to a 2D Schr\"odinger operator. The numerical scheme is applicable to a general 2D Schr\"odinger operator with fast decaying potential.
\end{abstract}

\section{Introduction}

This paper concerns numerical calculation of the discrete spectra of some  two-dimensional Schr\"odinger operators with soliton potential. These potentials were found in \cite{TT1,TT2} by using the Moutard transformation well-known in surface geometry
and have interesting spectral properties. The numerical scheme is based on the Galerkin method and can be applied to general
operators with fast-decaying potentials.

The Darboux transformation (the ladder method) allows us to construct the integrable one-dimensional Schr\"odinger operators
\begin{equation}
\label{1dim}
H = -\frac{d^2}{dx^2} + u(x),
\end{equation}
i.e. the operators whose spectra and eigenfunctions are explicitly described.
To each integrable operator $H$ and a solution $\omega$ of the equation $H\omega=0$ there corresponds the new integrable operator
$\widetilde{H}$ of the same form. Namely this correspondence sends each solution $\psi$ of the equation $H \psi=E\psi$ to a function $\varphi = A \psi$ such that
$\widetilde{H}\varphi  = E\varphi$, where $A$ is some first order differential operator independent of $\psi$. Note that this correspondence is invertible.

For instance, the quantum harmonic oscillator is exactly the operator for which this transformation results in a shift of the potential $u(x)$
by a constant. This observation was used
by Dirac for finding the spectrum and eigenfunctions of the operator \cite{Dirac}. The successive application of the Darboux transformation, starting
with the trivial potential $u=0$, yields the rational solitons $\frac{n(n+1)}{x^2}$.

In fact, the Darboux transformation was originally derived as a one-dimen\-sio\-nal reduction of the Moutard transformation which
acts on two-dimensional Schr\"o\-din\-ger operators as follows:
\begin{equation}
\label{2dim}
H = -4\partial\bar{\partial} + u(x,y) = -\frac{\partial^2}{\partial x^2} - \frac{\partial^2}{\partial y^2} + u(x,y).
\end{equation}
Furthermore to each operator $H$ and a solution $\omega$ of the equation $H\omega=0$ there corresponds another operator $\widetilde{H}$ of this form,
which gives an explicit procedure of constructing solutions to the equation $\widetilde{H}\varphi=0$ from solutions of the equation $H\psi=0$.
The correspondence is one-to-one modulo $\frac{\mathrm{const}}{\omega}$.
In contrast to the Darboux transformation this method
gives no information on solutions of the equation
$$
\widetilde{H}\varphi  = E \varphi
$$
with $E \neq 0$.

For the potentials derived by the Moutard transformation it is interesting to know something beyond the
``zero energy level'' $E=0$. In this article we demonstrate how to achieve this aim numerically:

\begin{itemize}
\item
we calculate the discrete spectra of the Schr\"odinger operators with nontrivial kernel which were found in \cite{TT1};

\item
we describe the dynamics of the discrete spectra of the operators whose potentials depend on $t$, the temporary variable,
and give a blowing-up solution of the Novikov--Veselov equation \cite{TT2}.
\end{itemize}

We think that the numerical approach can be rather helpful for understanding the spectral problems,
will lead to formulating mathematical statements that are based on experiments and can be valuable for further proofs.
Some of these conclusions are presented as final Remarks.

\section{The Moutard transformation and some of its applications}

\subsection{The transformation}

Let $H$ be of the form (\ref{2dim})
and let $\omega$ be a solution to the equation
$$
H \omega = (- \Delta + u )\omega = 0.
$$
Then the Moutard transformation of $H$
is defined as
$$
\widetilde{H} = -\Delta + u -2\Delta \log \omega =
-\Delta - u + 2\frac{\omega_x^2+\omega_y^2}{\omega^2}.
$$

Straightforward computations show that
if $\psi$ satisfies the equation
$H \psi = 0$,
then the function $\varphi$ defined from the system
\begin{equation}
\label{eigen-moutard}
(\omega \varphi)_x = -\omega^2 \left(\frac{\psi}{\omega}\right)_y, \ \ \
(\omega \varphi)_y = \omega^2 \left(\frac{\psi}{\omega}\right)_x
\end{equation}
satisfies the equation
$$
\widetilde{H}\varphi = 0.
$$
If $\varphi$ meets (\ref{eigen-moutard}), then
$$
\varphi + \frac{C}{\omega}
$$
satisfies the same equation for every constant  $C$.

If the potential $u=u(x)$ depends only on $x$  and
$\omega=f(x)e^{\sqrt{E}y}$, then
$$
H_0 f = \left( - \frac{d^2}{dx^2} + u\right) f = Ef
$$
and the Moutard transformation reduces to the Darboux transformation
of $H_0$ defined by $f$:
$$
H = H_0  - \frac{\partial^2}{\partial y^2} \ \ \longrightarrow \ \
\widetilde{H} =
\widetilde{H_0} -\frac{\partial^2}{\partial y^2}.
$$

Originally, the Moutard transformation was introduced for the hyperbolic operators
$$
\partial_\xi \partial_\eta + u,
$$
and was used, for instance, for constructing new negatively curved surfaces in $\R^3$ from those already available.
However, since the procedure is formally analytical we may put $\xi = z$ and $\eta = \bar{z}$ and derive its version for
Schr\"odinger operators.

\subsection{Two-dimensional Schr\"odinger operators with nontrivial kernel}
\label{2-2}

The Moutard transformation applied to the trivial potential $u=0$ does not give a nonsingular fast decaying potential. However such a potential for which
the scattering operator is well defined may be achieved by the double iteration of the Moutard transformation.
That was done in \cite{TT1} whose main results are as follows:

\begin{itemize}
\item
the potential (see Fig. \ref{figure:u1})
\begin{equation}
\label{p1}
u = -\frac{5120    (1 + 8    x + 2y + 17 x^2 + 17
y^2)}{(160 + 4(x^2 + y^2)(1 + 4x +y) + 17(x^2+y^2)^2)^2} =
\end{equation}
$$
-\frac{5120|1+(4-i)z|^2}{(160+|z|^2|2+(4-i)z|^2)^2}
$$
\begin{figure}[h]
\begin{center}
\includegraphics[scale=0.8]{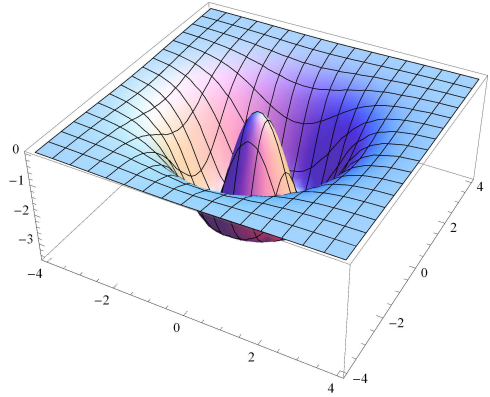}
\end{center}
\caption{The potential $u$ of the form(\ref{p1})}
\label{figure:u1}
\end{figure}
is nonsingular and decays as $r^{-6}$, while
the corresponding Schr\"odinger operator $H$ of the form (\ref{2dim})
has at least two-dimensional kernel (see Fig. \ref{figure:case1}):
$$
H\Psi_1 = H\Psi_2 = 0
$$
with
\begin{equation}
\label{p1f}
\begin{split}
\Psi_1 = \frac{x + 2    x^2 + x    y - 2    y^2}{160 + 4    x^2 +
4y^2 +  16    x^3  + 4    x^2    y + 16    x    y^2 + 4
y^3 + 17 (x^2+y^2)^2}, \\
\Psi_2 = \frac{2    x + 2y + 3    x^2 + 10    x    y - 3    y^2}{160
+ 4    x^2 + 4y^2 +  16    x^3  + 4    x^2    y + 16    x    y^2 + 4
y^3 + 17 (x^2+y^2)^2};
\end{split}
\end{equation}

\begin{figure}[h]
\begin{subfigure}[h]{0.5\linewidth}
\includegraphics[scale=0.5]{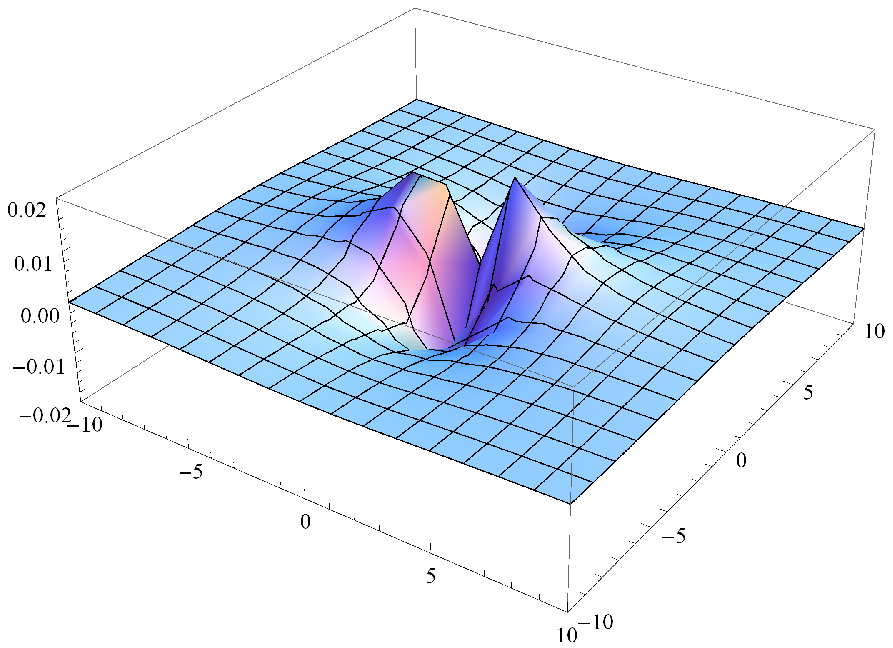}
\caption{$\Psi_1$}
\label{figure:psi11}
\end{subfigure}
\begin{subfigure}[h]{0.5\linewidth}
\includegraphics[scale=0.5]{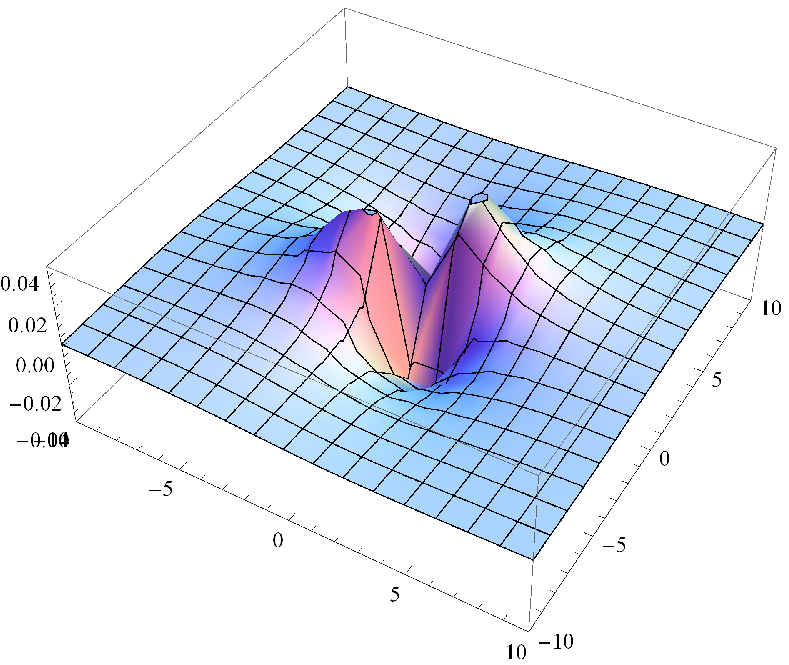}
\caption{$\Psi_2$}
\label{figure:psi12}
\end{subfigure}
\caption{The eigenfunctions $\Psi_1$ and $\Psi_2$ of the form (\ref{p1f})}
\label{figure:case1}
\end{figure}

\item
the potential
\begin{equation}
\label{p2}
u = \frac{F_0(x,y)}{G(x,y)^2}
\end{equation}
where
$$
F_0(x,y) = -1280000(25+20x-287x^2+60x^3+1800x^4 -30y -600xy -
$$
$$
300x^2y +
313y^2 +60xy^2 + 3600x^2y^2 -300y^3 + 1800y^4),
$$
$$
G(x,y) = 40000 + 100x^2 +40x^3 -387x^4 + 40x^5 +800x^6 -60x^2 y -
$$
$$
800x^3y -200 x^4y + 100y^2 + 40xy^2 + 26x^2y^2 + 80 x^3y^2 + 2400x^4y^2
-60y^3 -
$$
$$
800xy^3 -
400x^2y^3+  413y^4 + 40 xy^4 + 2400x^2y^4 -200y^5 + 800y^6,
$$
\noindent
is nonsingular and decays as $r^{-8}$, while the following functions $\psi_1$ and $\psi_2$
span
a two-dimensional subspace in the kernel of the Schr\"odinger operator:
$$
H\Psi_1 = H\Psi_2 = 0
$$
for
\begin{equation}
\label{p2f}
\Psi_1 = \frac{F_1(x,y)}{G(x,y)}, \ \ \
\Psi_2 = \frac{F_2(x,y)}{G(x,y)}
\end{equation}
where
$$
F_1(x,y) = -10x -2x^2 + 20 x^3 + 6xy + 60 x^2 y + 2y^2 - 60 x
y^2 - 20 y^3,
$$
$$
F_2(x,y) =
-10 x - 5 x^2 - 10 y + 2 x y + 120 x^2 y + 5 y^2 - 40
y^3.
$$
\end{itemize}

{\sc Remark.} Note that

1) some good scattering theory is available for an $N$-dimensional Schr\"o\-din\-ger operator if the potential of the operator
meets the condition
\begin{equation}
\label{scattering}
|u(x)| \leq \frac{\mathrm{const}}{(1+|x|)^{N+\varepsilon}}, \ \ x \in \R^N,
\end{equation}
with $\varepsilon >0$ an arbitrary positive constant \cite{Faddeev};

2) for the one-dimensional operators meeting the Faddeev condition
$$
\int_{-\infty}^{+\infty} |u(x)|(1+|x|)dx < \infty
$$
the discrete spectrum is finite and negative;

3)  for $N \geq 5$ it is easy to construct  a potential of the form $u(x) = \frac{\Delta \widetilde{G}}{\widetilde{G}}$, where $\widetilde{G}$ is a nonsingular positive function
coinciding with the fundamental solution of the Laplace equation $G(x) = \frac{\mathrm{const}}{|x|^{N-2}}$ outside some compact set, such that
$u(x)$ meets (\ref{scattering}) and the corresponding operator has nontrivial kernel.

Potentials (\ref{p1}) and (\ref{p2}) are the first examples
of the two-dimensional Schr\"odinger operators meeting (\ref{scattering}) and having a nontrivial
kernel. Similar examples in dimensions $N=3$ and $N=4$ have been unknown until recently.

\subsection{Blowing up solutions of the Novikov--Veselov equation}
\label{2-3}

The Novikov--Veselov (NV) equation \cite{NV} has the form
\begin{equation}
\label{nv}
\begin{split}
U_t = \partial^3 U + \bar{\partial}^3 U + 3\partial(VU) +
3\bar{\partial} (\bar{V}U) =0,
\\
\bar{\partial}V = \partial U,
\end{split}
\end{equation}
and can be derived as the compatibility condition of the system
\begin{equation}
\label{eq}
\begin{split}
H \varphi = 0, \\
\partial_t \varphi = (\partial^3 + \bar{\partial}^3 + 3V\partial +
3\bar{V}\bar{\partial})\varphi
\end{split}
\end{equation}
where
$\bar{\partial}V = \partial U, \partial \bar{V} = \bar{\partial} U$,
and
$$
H = \partial\bar{\partial} + U = \frac{1}{4}\Delta - \frac{u}{4}
$$
is a two-dimensional Schr\"odinger operator.

Its one-dimensional reduction corresponding to
$U=U(x)$ and $U=V$
is the Korteweg--de Vries equation
$$
U_t = \frac{1}{4}U_{xxx} + 6UU_x.
$$
The KdV equation admits the Lax representation
$$
\frac{dL}{dt} = [L,A],
$$
where $L$ the one-dimensional Schr\"odinger operator with potential $U$. Therefore, the KdV evolution preserves the whole spectrum of
$L$. In contrast to the KdV case, the NV equation has the Manakov triple representation
$$
\frac{dH}{dt} = [H,A] + BH
$$
and preserves  only the zero-level spectrum.

The Moutard transformation of $H$ can be extended to some transformation of solutions of the
NV equation to another solutions of the same equation.
By using the transformation, in \cite{TT2} the first blowing up solution of the NV equation was constructed which is as follows:
\begin{equation}
\label{blowup}
\begin{split}
U= 2\partial\bar{\partial} \log \Phi, \ \ \ V = 2\partial^2 \log \Phi,
\\
\Phi(x,y,t) = 3(x^2+y^2) + 4(x^3+y^3)+30-12t.
\end{split}
\end{equation}

\begin{figure}[h]
\begin{subfigure}[h]{0.5\linewidth}
\includegraphics[scale=0.4]{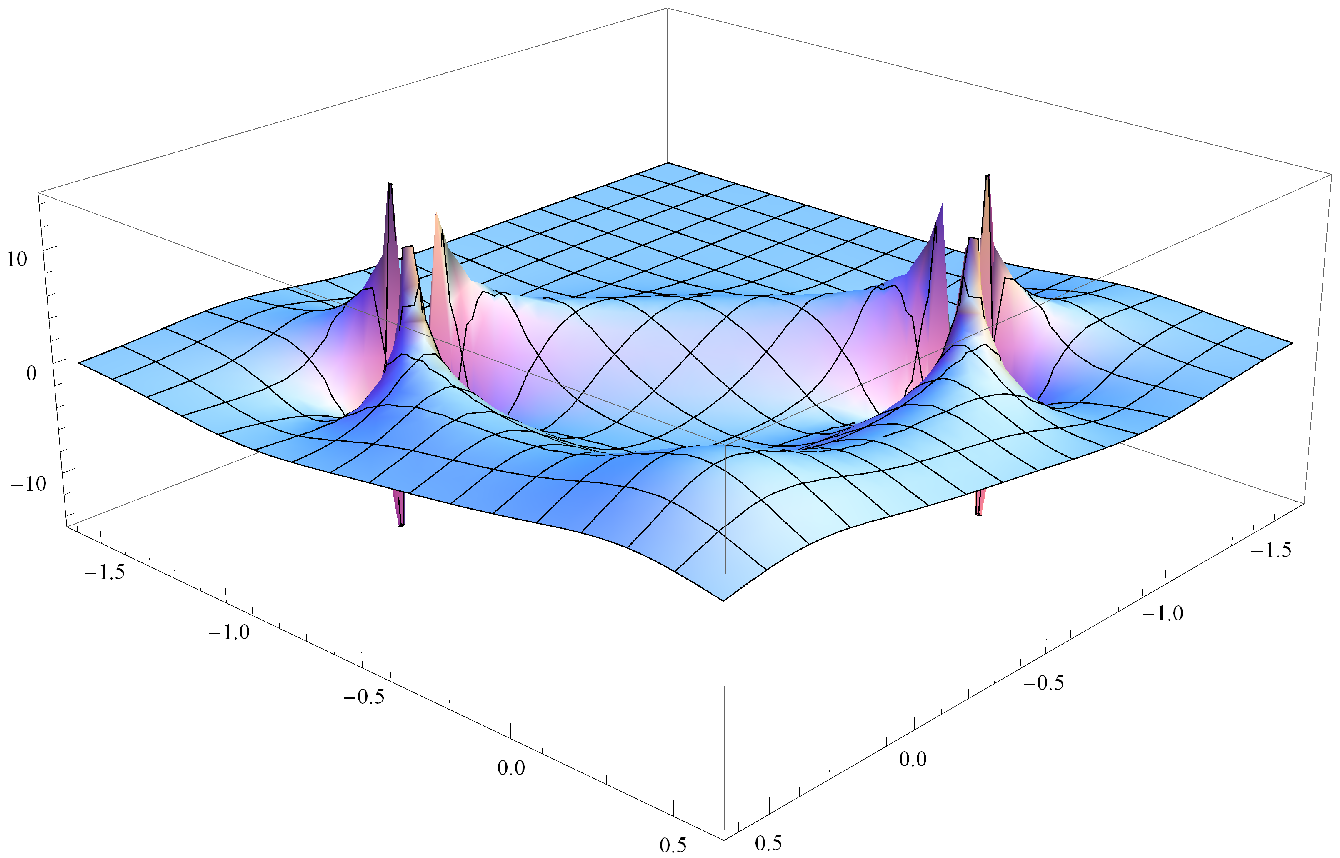}
\caption{$t \to t_\ast$}
\label{figure:blowup1}
\end{subfigure}
\begin{subfigure}[h]{0.5\linewidth}
\includegraphics[scale=0.4]{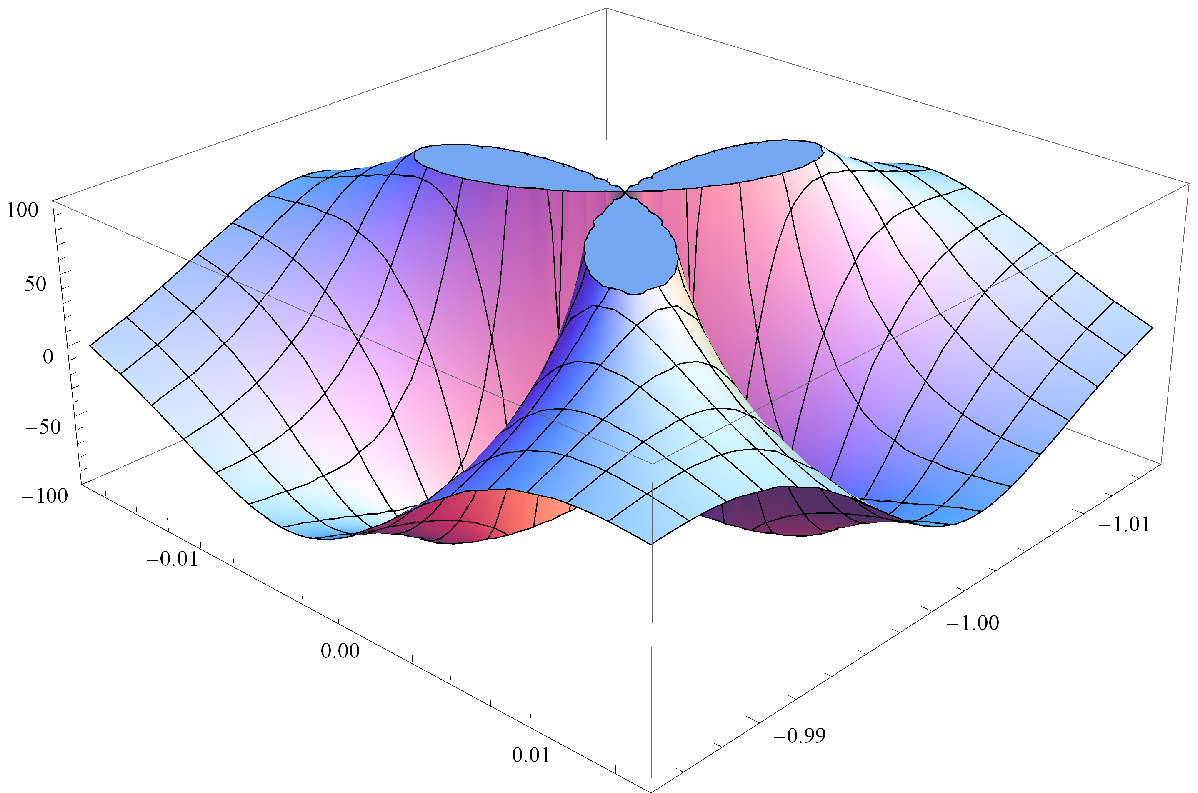}
\caption{A singular point}
\label{figure:blowup2}
\end{subfigure}
\caption{The singularities of $U$}
\label{figure:blowup}
\end{figure}

For $t< t_\ast$ the potential $U$ is nonsingular and decays as $r^{-3}$.
As $t$ approaches $t_\ast$ from below, the solution starts to oscillate in a bounded domain
and at $t=t_\ast= \frac{29}{12}$
the function $\Phi$ vanishes at a couple of points at which $U$ has singularities:
$U$ is a rational function and for $t=t_\ast$ the denominator vanishes
at these points, but the nominator vanishes at certain lines passing through the same points (see Fig. \ref{figure:blowup}).

We recall that
$I=\int_{\R^2} U\, dx\,dy$ is the first integral of the NV equation, which restricts the possible types of singularities.

{\sc Remarks.}
1) For the KdV equation, i.e. the one-dimensional reduction of the NV equation, such a blowup is impossible. For instance,
a solution of the KdV equation with smooth initial data $u_0(x) = o(|x|)$ as $|x| \to \infty$ is regular and unique for all times $t$ \cite{Menikoff}.

2) In \cite{KM} it was noted that the more general function
$U = 2\partial\bar{\partial} \log (a(x^2+y^2)^2 + b(x^3+y^3)+c-4bt)$, with $a,b$ and $c$ real constants, meets the NV equation and gives a similar blowing up solution.

3) The modified Novikov--Veselov (mNV) equation, which is a similar two-dimensional version of the modified Korteweg--de Vries equation,
also admits blowups which are of a different nature. For this equation $I= \int U^2\,dx\,dy$ is the first integral in contrast to the NV equation
whose evolution preserves $\int U\, dx\,dy$. Recently in \cite{T1,T2} there was constructed a solution to the mNV equation on the whole plane
which decays rather fast at infinity and is real-analytic outside the only singular point $x=y=t=0$, while the first integral $I$ equals $3\pi$ for $t \neq 0$ and jumps to
$2\pi$ at $t=0$.

\section{A numerical method}

\subsection{The basic scheme}
\label{3-1}

We will briefly explain the scheme in general.

Take an orthonormal basis $\{e_k\}, k=1,2,\dots$, for $L_2(\R^m)$:
$$
\langle e_j | e_k \rangle = \delta_{jk}, j,k=1,\dots,
$$
where
$\langle u | v\rangle$
is the standard inner product on $L_2(\R^m)$.

Denote by
$P_n$ the orthogonal projection to the finite-dimensional subspace $V_k$ spanned by the first $k$ vectors:
$$
P_n: L_2 \to V_n = \C e_1 + \dots + \C e_n.
$$
Given an operator $H: L_2(\R^m) \to L_2(\R^m)$ we consider its approximation of the form
$$
H_n = P_n H: V_n \to V_n
$$
and find the spectrum of $H_n$:
$$
E^{(n)} = \{E^{(n)}_1 \leq  \dots \leq E^{(n)}_n\}.
$$
The operator $H_n$ is represented in the basis $e_1,\dots,e_n$ by the matrix
$$
A_{jk} = \langle e_j | H_n e_k \rangle = \langle e_j | H e_k \rangle, \ \ 1 \leq j,k \leq n,
$$
which is the principal minor of the matrix $A$ describing operator $H$ in the basis $\{e_k\}$.

Let $H$ be the Schr\"odinger operator
$$
H = -\left(\frac{\partial^2}{\partial x^2} +\frac{\partial^2}{\partial y^2}\right) + u(x,y)
$$
with a fast decaying potential.
The number of eigenvalues of $H_n$ counted with multiplicities is equal to $n$ and grows as
$n \to \infty$. Hence the whole spectrum $E^{(n)}$ cannot be considered as approximation of
the discrete spectrum of $H$ which can be even empty.

However it is known that if the potential decays rather fast, i.e.,
$$
u = o\left(\frac{1}{|x|}\right)  \ \ \ \mbox{as $|x| \to \infty$};
$$
then the Schr\"odinger operator has no positive eigenvalues \cite{Kato}.
Hence

{\sl the sequence $E^{(n)}$ with the excluded positive elements not very close to zero, i.e., the truncated sequence,
can be considered as approximation of the discrete spectrum of $H$ as $n \to \infty$.}

We do not supply any strong mathematical evidence to this claim. But similar arguments are widely used in
computational chemistry (quantum chemistry and molecular dynamics) and the results fit experiments.
In this article we present the results of a similar numerical study of the discrete spectra of potentials from \ref{2-2} and {\ref{2-3}.

\subsection{A numerical scheme based on the Hermite functions}

The Hermite functions
$$
\varphi_k = \frac{(-1)^k}{\sqrt{2^k k! \sqrt{\pi}}}e^{\frac{x^2}{2}} \frac{d^k}{dx^k} e^{-x^2}, \ \ k=0,1,2,\dots,
$$
form the orthonormal basis for $L_2(\R)$, i.e.
$$
\langle \varphi_j | \varphi_k \rangle = \delta_{jk},
$$
and are the eigenfunctions of the quantum harmonic oscillator
$$
\left(-\frac{d^2}{dx^2} + x^2\right) \varphi_k = (2k+1)\varphi_k.
$$

Let us consider in $L_2(\R^2)$ the orthogonal basis formed by the products
$$
\varphi_{j,k}(x,y) = \varphi_j(x) \varphi_k(y)
$$
of the Hermite functions of $x$ and $y$.
In this basis the Schr\"odinger operator is described by the matrix
$$
A_{(j,k)(l,m)} = \langle \varphi_{j,k} | H \varphi_{l,m} \rangle.
$$
Since $\psi_{j,k}$ is the eigenfunction of the two-dimensional quantum harmonic oscillator
$$
(-\Delta + x^2+y^2) \varphi_{j,k} = \left(-\frac{\partial^2}{\partial x^2} - \frac{\partial^2}{\partial y^2} + x^2 + y^2\right)\varphi_{j,k} =
2(j+k+1)\varphi_{j,k},
$$
we have
$$
\langle \varphi_{j,k} | H \varphi_{l,m} \rangle = \langle \varphi_{j,k} | (-\Delta+u) \varphi_{l,m}\rangle =
\langle \varphi_{j,k} | (u-x^2-y^2 + 2(l+m+1)) \varphi_{l,m}\rangle =
$$
$$
\langle \varphi_{j,k} | (u-x^2-y^2) \varphi_{l,m}\rangle + 2(j+k+1)\delta_{jl}\delta_{km}.
$$

In the sequel let
$H_{(N)}$ denote the operator
$$
H_{(N)} = P_{(N)} H P_{(N)}
$$
where
$$
P_{(N)}: L_2(\R^2) \to \mathrm{span}\,\{\varphi_{j,k}, 0 \leq j,k \leq N-1\}
$$
is the orthogonal projection of $L_2(\R^2)$ to the $N^2$-dimensional subspace spanned by the two-dimensional Hermite functions
$\varphi_{j,k}$ with $0 \leq j,k \leq N-1$.

\subsection{The Gauss-Hermite quadrature for matrix elements }

For calculating the matrix elements $A_{(j,k)(l,m)}$ we use the Gauss--Hermite quadrature (see \cite{Lubich}):
the integral over $\mathbb{R}$ of a function of the form
$$
f(x) = e^{-x^2}h_s(x),
$$
with $h_s(x)$ a polynomial of degree $s$, is evaluated by the quadrature formula
$$
\int\limits_{-\infty}^{\infty}f(x)dx = \int\limits_{-\infty}^{\infty}e^{-x^2}h(x)dx =
\sum\limits_{i=1}^{n}w_i h_s(x) =
\sum\limits_{i=1}^{n}\omega_i f(x_i),
$$
where $x_i$ are the roots of the Hermite polynomial of degree $n$ while $w_i, \;\omega_i$ are the corresponding weights:
$$
\omega_i = \frac{2^{n-1}n!\sqrt{\pi}}{n^2(H_{n-1}(x_i))^2}e^{x_i^2}, \quad  w_i = \frac{2^{n-1}n!\sqrt{\pi}}{n^2(H_{n-1}(x_i))^2}
$$
This quadrature is exact for all $s$ up to $2n-1$. If $f(x)=g(x)e^{-x^2/2}$ with $g(x)\in L^2(\mathbb{R})$ for which the coefficients $c_k=(\phi_k|g)$ if
the Hermite expansion of $g$ satisfy $|c_k|\leq C(1+k)^{-r}$, then the quadrature error is bounded by $O(n^{-r})$.

In the two-dimensional case
$$
\int\limits_{-\infty}^{\infty}\int\limits_{-\infty}^{\infty}f(x,y)dxdy =
\sum\limits_{i=1}^{n}\sum\limits_{j=1}^{n}\omega_i \omega_j f(x_i, y_j).
$$

For calculating the roots of the Hermite polynomials  we use the algorithm that is proposed in \cite{Glaser}.

\section{Numerical results}

\subsection{The Schr\"odinger operators with nontrivial kernel}

{\bf 1.} The potential written down in (\ref{p1}) is nonpositive and vanishes exactly at one point. Therefore, by the Rayleigh principle, the discrete spectrum
of the corresponding Schr\"odinger operator $H$ is nontrivial
and the first eigenvalue equals to
$$
\lambda_1 = \inf_{|\psi|=1} \langle \psi | H \psi \rangle.
$$

It appears that except the five eigenvalues $\lambda_1,\lambda_2,\lambda_3,\lambda_{01}$ and $\lambda_{02}$
the eigenvalues of $H_{(N)}$ are greater than $0.1$ for $N =  16, 20, 25, 32$, and $50$.
The exceptional five eigenvalues split into the two groups:

a) the negative eigenvalues $\lambda_1, \lambda_2,\lambda_3$ that are separated from $0$;

b) the positive eigenvalues $\lambda_{01}$ and $\lambda_{02}$ that converge to the zero from above as $N$ grows.

The results of computations are given in Table \ref{table:lambdas1}.

\begin{table}[h!]
\begin{center}
\begin{tabular}{l|c|c|c|c|c}
$\lambda$ & $N^2=256$ & $N^2 = 400$ & $N^2 = 625$ &$N^2 = 1024$ & $N^2=2500$ \\
\hline
$\lambda_1$ & $-1.80934$ & $-1.8093$ & $-1.80935$ & $-1.80936$ & $-1.80936$\\
$\lambda_2$ & $-1.09104$ & $-1.09134$ & $-1.09158$ & $-1.09163$ & $-1.09163$\\
$\lambda_3$ & $-1.01904$ & $-1.01929$ & $-1.01927$ & $-1.01927$ & $-1.01927$\\
$\lambda_{01}$&  $0.0205608$ &  $0.0129299$& $0.00811659$& $0.00487366$ & $0.00194779$\\
$\lambda_{02}$&  $0.0320718$ & $0.0199062$  & $0.0125$ & $0.00745718$ & $0.0029613$ \\

\end{tabular}\\
\caption{The approximate eigenvalues of the Schr\"odinger operator with the soliton potential (\ref{p1})}
\label{table:lambdas1}
\end{center}
\end{table}

Denote by $W$ the span of the eigenfunctions $\Psi_1$ and $\Psi_2$ of the form (\ref{p1f}) and by $\psi_1$, $\psi_2$, $\psi_3$, $\psi_{01}$, $\psi_{02}$
the eigenvectors of $H_{(N)}$ corresponding to $\lambda_1$, $\lambda_2$, $\lambda_3$, $\lambda_{01}$, $\lambda_{02}$. Let $\alpha_i$ stand for the angle between
$\psi_{0i}$ and $W$, $i=1,2$.  The calculations in Table \ref{table:coslambda1} show that $\psi_{01}$ and $\psi_{02}$ converge to functions in $W$ as $N$ grows.

Therefore we arrive at the following:

\begin{conclusion}
The discrete spectrum of the operator $H$ with potential (\ref{p1}) consists of five eigenvalues (their approximations are given in Table 1)
of which three are negative.
The kernel of $H$ is two-dimensional.
\end{conclusion}

\begin{table}[h!]
\begin{center}
\begin{tabular}{l|c|c|c|c|c|c|c}
$\cos \alpha$ &$N^2 = 256$ &$N^2=400$ & $N^2=625$ & $N^2=1024$  & $N^2=2500$ \\
\hline
$\cos \alpha_1 $ & $0.991287$ & $0.992226$ & $0.993199$ & $0.994215$ & $0.995846$\\
$\cos \alpha_2 $ & $0.991362$ & $0.992084$ & $0.992921$ & $0.993901$ & $0.995526$ \\
\end{tabular}\\
\end{center}
\caption{The cosines of the angles between the eigenvectors corresponding to $\lambda_{01}$ and $\lambda_{02}$ and
the linear span of $\Psi_1$ and $\Psi_2$}
\label{table:coslambda1}
\end{table}

The fast decay of the coefficients of expansions of $\psi_\alpha$ in $\varphi_{i,j},\;(i,j=0,...,N-1)$ is demonstrated
by the graphs of these coefficients parameterized by $i,j$ and calculated for $N=32$ (see Fig. \ref{figure:eigenvectors1}).
The graphs of these eigenvectors are presented in Fig. \ref{figure:eigenfunctions1}.

\begin{figure}[h!]
\begin{center}
\begin{subfigure}[h!]{0.33\linewidth}
\includegraphics[scale=0.6]{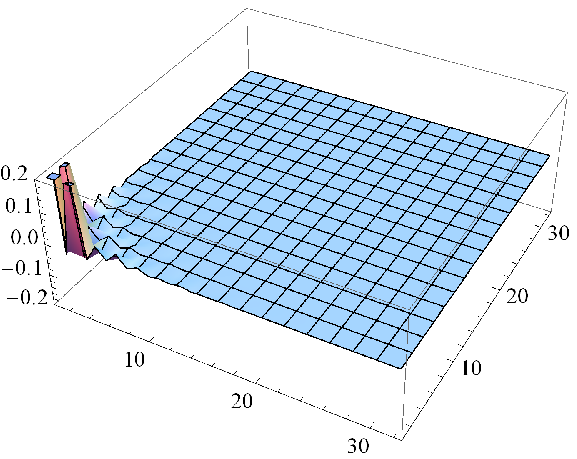}
\caption{$\psi_1$}
\label{figure:S11}
\end{subfigure}
\begin{subfigure}[h!]{0.3\linewidth}
\includegraphics[scale=0.6]{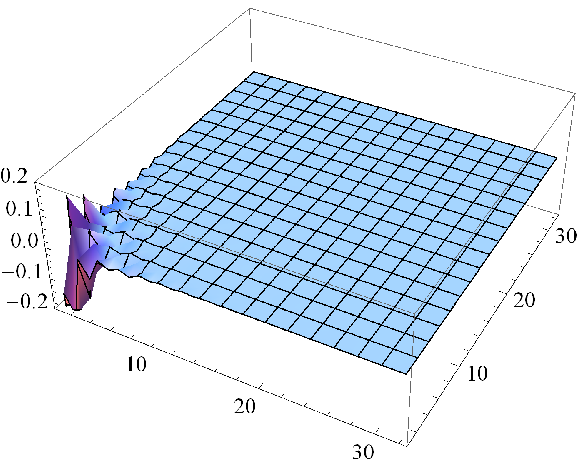}
\caption{$\psi_2$}
\label{figure:S12}
\end{subfigure}
\begin{subfigure}[h!]{0.3\linewidth}
\includegraphics[scale=0.6]{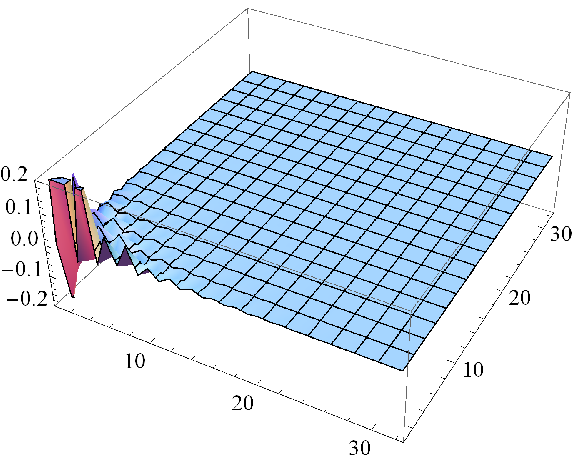}
\caption{$\psi_3$}
\label{figure:S13}
\end{subfigure}
\\
\begin{subfigure}[h!]{0.3\linewidth}
\includegraphics[scale=0.6]{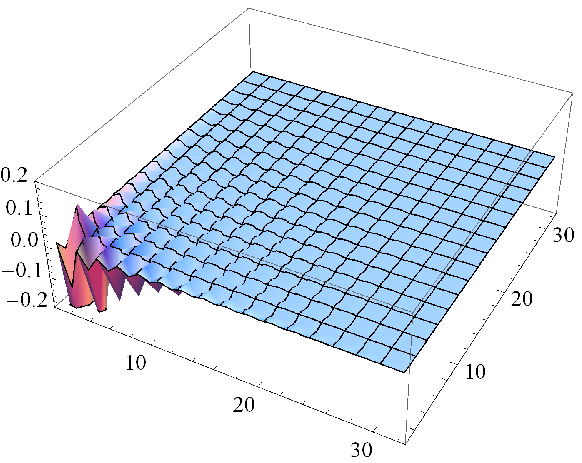}
\caption{$\psi_{01}$}
\label{figure:S101}
\end{subfigure}
\begin{subfigure}[h!]{0.3\linewidth}
\includegraphics[scale=0.6]{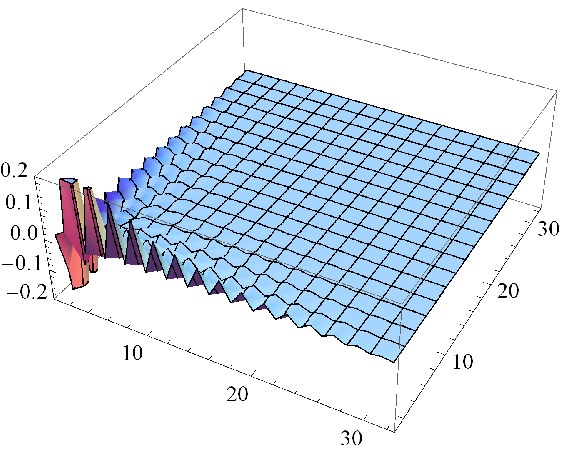}
\caption{$\psi_{02}$}
\label{figure:S102}
\end{subfigure}
\caption{The expansions of eigenvectors}
\label{figure:eigenvectors1}
\end{center}
\end{figure}

\begin{figure}[h!]
\begin{center}
\begin{subfigure}[h!]{0.33\linewidth}
\includegraphics[scale=0.6]{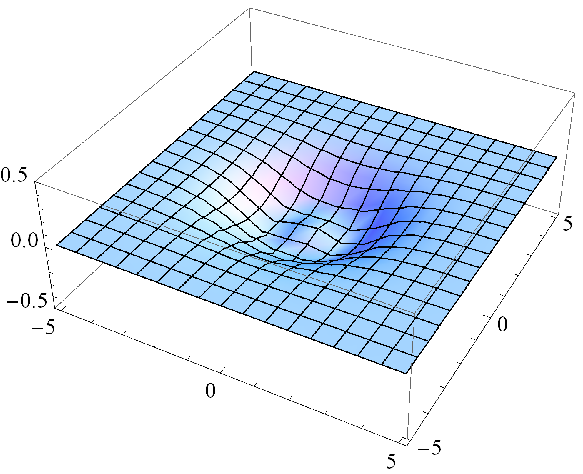}
\caption{$\psi_1$}
\label{figure:p11}
\end{subfigure}
\begin{subfigure}[h!]{0.3\linewidth}
\includegraphics[scale=0.6]{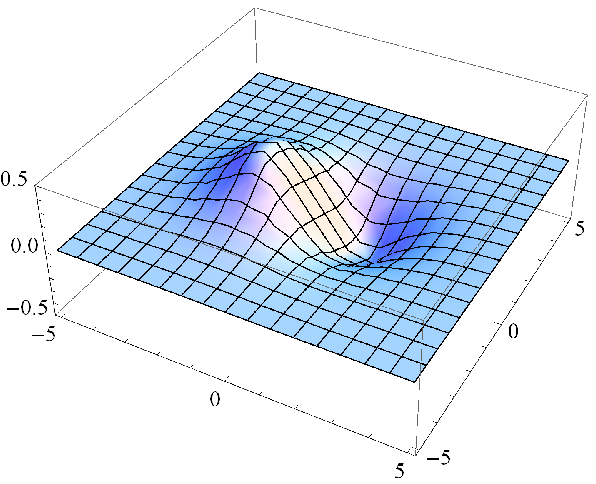}
\caption{$\psi_2$}
\label{figure:p12}
\end{subfigure}
\begin{subfigure}[h!]{0.3\linewidth}
\includegraphics[scale=0.6]{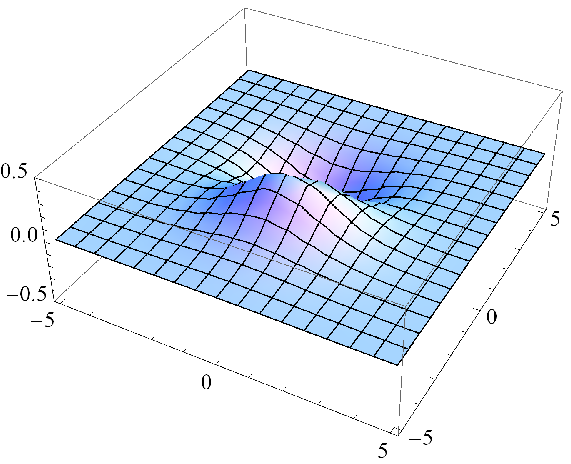}
\caption{$\psi_3$}
\label{figure:p13}
\end{subfigure}
\\
\begin{subfigure}[h!]{0.3\linewidth}
\includegraphics[scale=0.6]{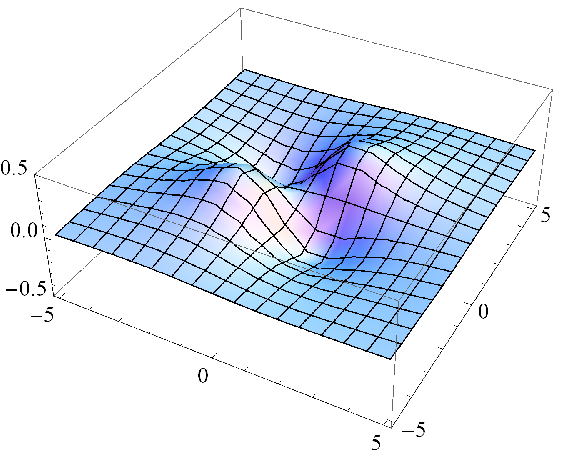}
\caption{$\psi_{01}$}
\label{figure:p14}
\end{subfigure}
\begin{subfigure}[h!]{0.3\linewidth}
\includegraphics[scale=0.6]{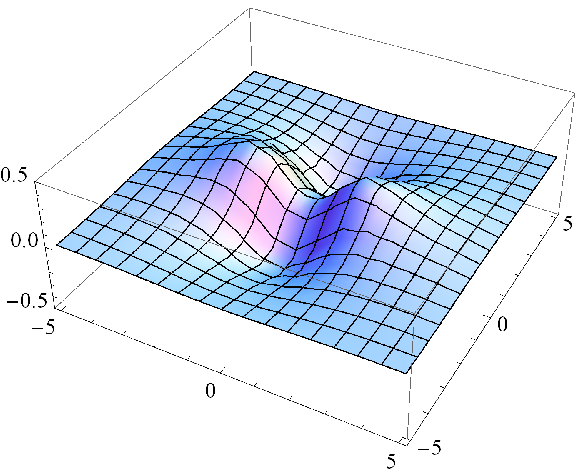}
\caption{$\psi_{02}$}
\label{figure:p15}
\end{subfigure}
\caption{The eigenvectors}
\label{figure:eigenfunctions1}
\end{center}
\end{figure}

\newpage
{\bf 2.} The analogous calculations for potential (\ref{p2}), which is also nonpositive, shows
that the eigenvalues of $H_{(N)}$ that may correspond to the eigenvalues of $H$
split into the two groups:

a) the negative eigenvalues $\lambda_1,\dots,\lambda_5$ separated from zero;

b) the eigenvalues $\lambda_{01}, \lambda_{02}$ converging to zero as $N$ grows.

The calculations of eigenvalues are presented in Table \ref{table:lambdas2}.

\begin{table}[h!]
\begin{center}
\begin{tabular}{l|c|c|c|c|c}
$\lambda$ & $N^2=256$ & $N^2 = 400$ & $N^2 = 625$ &$N^2 = 1024$ & $N^2=2500$ \\
\hline
$\lambda_1$ & $-2.83707$ & $	-2.8501$ & $	-2.84788$ & $-2.84807$ & $-2.84812$ \\
$\lambda_2$ & $-2.49391$ & $	-2.51335$ & $-2.50945$ & $-2.50925$ & $-2.50919$ \\
$\lambda_3$ & $-2.23477$ & $	-2.2435$ & $	-2.24252$ & $-2.24291$ & $-2.24302$ \\
$\lambda_4$ & $-1.33696$ & $	-1.34495$ & $-1.33922$ & $-1.33888$ & $-1.33882$ \\
$\lambda_5$ & $-1.31464$ & $	-1.33525	$ & $-1.33505$ & $-1.33493$ & $-1.33487$ \\
$\lambda_{01}$ & $0.008239$ & $	0.000002$ & $0.000656$ & $0.000626$ & $0.000237072$ \\
$\lambda_{02}$ & $0.019057$ & $	0.001244$ & $0.001943$ & $0.000966$ & $0.000248573$ \\
\end{tabular}\\
\caption{The approximate eigenvalues of the Schr\"odinger operator with the soliton potential (\ref{p2})}
\label{table:lambdas2}
\end{center}
\end{table}

The eigenvectors $\psi_{01}$ and $\psi_{02}$ corresponding to $\lambda_{01}$ and $\lambda_{02}$
have to approximate functions from the kernel. The cosines of the angles $\alpha_i, i=1,2$,
between the eigenvectors corresponding to $\lambda_{01}$ and $\lambda_{02}$ and
the linear span of $\Psi_1$ and $\Psi_2$ are presented in Table \ref{table:coslambda2}

\begin{table}[h!]
\begin{center}
\begin{tabular}{l|c|c|c|c|c|c|c}
$\cos \alpha$ &$N^2 = 256$ &$N^2=400$ & $N^2=625$ & $N^2=1024$ & $N^2=2500$\\
\hline
$\cos \alpha_1$ & $0.998691$ & $0.998852$ & $0.999245$ & $0.999499$ & $0.999764$ \\
$\cos \alpha_2$ & $0.998792$ & $0.998910$ & $0.999273$ & $0.999505$ & $0.999765$ \\
\end{tabular}\\
\end{center}
\caption{The cosines of the angles between the eigenvectors corresponding to $\lambda_{01}$ and $\lambda_{02}$ and
the linear span of $\Psi_1$ and $\Psi_2$}
\label{table:coslambda2}
\end{table}

Therefore, we arrived at

\begin{conclusion}
The discrete spectrum of the operator $H$ with potential (\ref{p2}) consists of seven eigenvalues
(their approximations are given in Table \ref{table:lambdas2}) of which five are negative.
The kernel of $H$ is two-dimensional.
\end{conclusion}

\subsection{The blowing up solution of the Novikov--Veselov equation}

For the operator
$$
H = -\Delta - 4U
$$
where $U$ is of the form (\ref{blowup}), i.e., a blowing up solution of the NV equation,  we numerically find the five eigenvalues:
$$
\lambda_1, \lambda_{2}, \lambda_{3}, \lambda_{01}, \lambda_{02}.
$$
At $t=0$ the three eigenvalues $\lambda_1, \lambda_{2}, \lambda_{3}$ are negative and lie rather far from zero, whereas
$\lambda_{01}$ and $\lambda_{02}$ are close to zero and approximate the
zero eigenvalue.
Indeed, the kernel is at least two-dimensional and includes the subspace that is spanned by
$$
\Psi_1(t,x,y)=\frac{2x^2-2y^2+2x+2y}{3x^4+4x^3+6x^2y^2+3y^4+4y^3+30-12t},
$$
$$
\Psi_2(t,x,y)=\frac{-4xy}{3x^4+4x^3+6x^2y^2+3y^4+4y^3+30-12t}.
$$

Since we have an explicit description of the evolution of the potential we can calculate the dynamics of the eigenvalues, which
is presented in Fig. \ref{Figure:VNL50x50}.

\begin{figure}[h!]
\begin{center}
\includegraphics[scale=0.6]{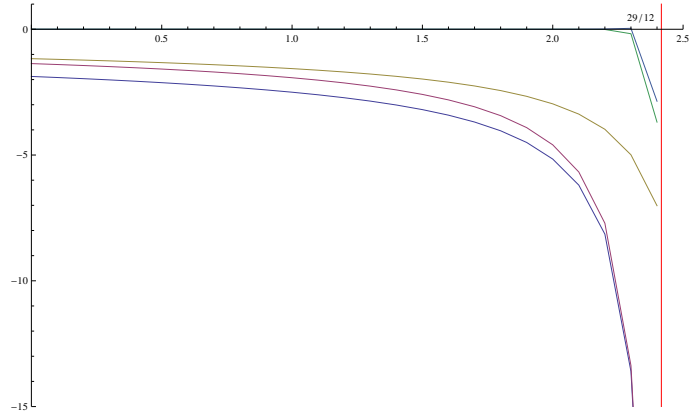}
\end{center}
\caption{The approximate dynamics of the eigenvalues for $N=50$}
\label{Figure:VNL50x50}
\end{figure}

The vertical line denotes the critical time $T_\ast=\frac{29}{12}$ at which the potential becomes singular.

We see the following:

\begin{itemize}
\item
Up to $t \approx 2.2$ the two maximal eigenvalues stay closed to the zero as it has to be because the zero-level spectrum is preserved
by this evolution. Therewith the negative spectrum evolves.

\item
After $t \approx 2.2$ the approximations to the zero eigenvalue substantially decrease going away from zero and therefore the calculations become unreliable.
The correct description of the evolution of the discrete spectrum for $t >2.2$ needs the Galerkin approximations with respect to higher-dimensional
subspaces or a more precise numerical scheme.
\end{itemize}

\section{Final remarks}

The calculations  were done by using the LAPACK package.

The main points for us consist in

1) demonstrating the possibilities of the method and we see, for instance, that the kernel is detected rather precisely;

2) studying by numerical methods the spectral properties of some two-dimen\-sio\-nal differential operators.

Usually the results of calculations are compared with an integrable case. Since the lack of examples of fast-decaying
two-dimensional potentials with explicitly described discrete spectrum we use for such a comparison the zero-level spectrum
explicitly established for the potentials that are considered in \cite{TT1}.

These calculations provide the numerical evidence for the following statements:

\begin{enumerate}
\item
{\sl The negative discrete spectrum is not preserved by the Novikov--Veselov flow and it seems that all conservation laws
are given by the spectral data of $H$ which correspond to the zero energy level (see \cite{TTF})}.

\item
The Darboux transformation can be interpreted as a transformation that adds or removes a point from the discrete spectrum
of a one-dimensional operator (\ref{1dim}). The operators from 2.2 and 2.3 are derived by the double iteration of the Moutard transformation starting at the zero potential $u=0$. The analogy between these transformations and the numerical calculations leads us to the conjecture that if the Moutard transformation respects the function classes of potentials and eigenfunctions then

{\sl the dimension of the kernel of $H$ is changed by one by the Moutard transformation and there are no such estimates related to the negative part of the discrete spectrum}.
\end{enumerate}

\medskip

{\bf Acknowledgement.}We thank C. Lubich  for stimulating discussions and, in particular, for the suggestion to apply the Galerkin method.

This work was supported by the Science Committee of the Ministry
of Education and Science of Republic of Kazakhstan
(program №0217/ptf-14-OT), RFBR (grant 15-01-01671) and
the Government of the Russian Federation (contract 14.Y26.31.0006).}

\end{document}